# PERSONALITY WIRELESS SENSOR NETWORKS (PWSN s)

**Muhammad Imran, Halabi B. Hasbullah, Abas Md Said**

Department of Computer & Information Sciences,
Universiti Teknologi PETRONAS,
Bandar Seri Iskandar, 31750 Tronoh, Perak, Malaysia.
*cmimran81@yahoo.com*, *halabi@petronas.com.my*, *abass@petronas.com.my*

**ABSTRACT**. *In recent years, WSNs are garnering lot of interest from re search community because of their unique characteristics and potential for enormous range of applications. Envision for new class of applications are being emerged such as human augmentation, enhancing social interaction etc. Misunderstanding or misinterpr etation of behaviors from individuals leads to social conflicts. There are various theories that classify people into different personality types. Most of the existing theories rely on questionnaires, which is highly unreliable. Anyone can lead such theori es in practice to incorrect classification intentionally or unintentionally. The objective of this research is to investigate existing solutions and propose a basic infrastructure for an automated context-aware psychological classification based on differe nt parameters. The idea is to use wearable sensors to sense and measure various human body parameters (i.e. body temperature, blood pressure, perspiration, brain impulses etc) that coerce human psychological condition. The data collected from these paramet ers is transformed in to information, to determine personality type, mood and psychological condition of interacting parties. This information is shared among counterparts to better understand each other in order to avoid potential conflicting situations. We believe that it will help peoples understand each other, improve their quality of life and minimize possible conflicting situations.*

## INTRODUCTION

Wireless sensor networks (WSNs) are getting world wide attention, In recent years, the rapid development in miniaturization; low power wireless communication, micro sensor, and microprocessor hardware have enable d the development of low-cost, low-power, multifunctional sensor nodes. These nodes are small in size and communicate in short distances. These tiny sensor nodes consisting of sensing, data processing and communication components leverage the idea of sensor networks based on collaborati ve effort of a number of nodes (Akyildiz *et al*., 2002).
A Wireless sensor network (WSN) is composed of large number of tiny low powered sensor nodes and one or multiple base stations (sinks). The sensor nodes sense, measure and collect ambient environment conditions and transform them into electric signal. These nodes use their processing abilities to carry out simple computations and send partially processed sensed data via the radio transmitter, to a base station either dir ectly or through a gateway. The gateway can perform fusion of the sensed data in order to filter out erroneous data and anomalies and to draw conclusions from the reported data over a period of time. A comprehensive overview of wireless sensor networks and their broad range of applications can be found in (Akyildiz *et al*., 2002, Yick et al., 2008 and Younis et al. 2004).

Large number of innovative set of applications for WSNs has been proposed in (Marron et al., 2006). It includes some new categories of applications like human augmentation and enhancing social interactions. Human augmentation refers to all the ubiquitous cooperating object technology that can be employed to assist our daily activities. Whereas enhancing social interactions uses cooperating objects to establish or maintain social relationships among people. The idea of PWSN falls under the category of enhancing social interactions.

Communication is most important aspect in building, maintaining and enhancing social interactions. While communication, personality type had a great influence on receiver. Personality type reveals important things about you, such as whether you're naturally more outgoing or reserved, realistic or imaginative, logical or sensitive, and organized or spontaneous (Falikowski, 2002 and Myers-Briggs). Interpretation is determined by attitudinal preference (introvert/extrovert); mode of information gathering (sensing/intuition); the way decisions are made (thinking/feeling); and how we orient (judging/perceiving) (Falikowski, 2002).

Misunderstanding or misinterpretation of behaviors from individuals leads to social conflicts. It effects people in their social interactions i.e. friends, families and organizations. Such situations lead to irritating decisions and may result in un-recoverable losses. Human nature is subject to change in different contexts.

Improvement in social interactions can reduce potential conflicts. It can be achieved by knowing one's own character and character of his counterpart in different contexts. It can be determined through one's personality type, mood and psychological condition. Most of the existing theories classify peoples into different personality types based on questionnaire, which is highly unreliable. The authors envisaged use of emerging technologies and presented their ideas in (Benenson et al., 2006 and Marron et al., 2006). They did not either consider technical details or they believe that such applications are far away to become reality.

The objective of this research is to investigate existing solutions and propose a basic infrastructure to determine an automated context-aware psychological classification in the form of personality wireless sensor networks.

Our proposed idea is to use tiny sensors (either wearable or implantable) into the human body to monitor vital body parameters. The person's character (personality type, mood and psychological condition) are determined in different contexts through these parameters. The person is notified with his own character and the character of counterpart in order to increase level of understanding.

The rest of the paper is organized as follows. Section II presents critical review of related work. In section III, we have described the problem statement. In section IV, we have presented our proposed infrastructure required for PWSN. Section V presents the application scenario as case study and section VI presents the conclusion of this paper. We have foreseen some future directions and briefly describe them in section VII.

# PROBLEM STATEMENT

In all times, misunderstandings among individuals and social groups lead to conflicts. It effects people in their social interactions i.e. friends, families and organizations. The closer friends become alien because of misinterpretation. Misunderstandings among life partners can lead to divorces, misinterpretations of behavior of the parents or children may lead to drifting apart of families. The performance of organizations largely depends on attitude of individuals working in different teams (Benenson et al., 2006). Similarly societies suffer from it because of chain reaction.

These situations lead to irritating decisions such as in-appropriate team formation, separating families apart that may result in un-recoverable losses. Human nature is subject to change in different contexts. They behave differently in different communities and situations. So there is a need to determine context-aware psychological classification.

The objective of this research is to investigate existing solutions and propose a basic infrastructure to determine an automated context-aware psychological classification in the form of personality wireless sensor networks.

# RELATED WORK

The improvement in social relationships can reduce potential conflicts and can help each and every person in their daily life. This can be achieved if the interacting parties are aware of their own characters and the characters of the counterparts. Of course, all people are different (Benenson et al., 2006). To the best of our knowledge and literature review, there are various techniques based on (Myers-Briggs) that classify people according to their personality types. One of these systems was invented by Briggs and Myers in early twentieth century and was successfully adopted by organizations in forming project teams. There are various challenges in applying psychological classifications. Most of the existing theories rely on questionnaires, which is highly unreliable. Anyone can lead such theories in practice to incorrect classification intentionally or unintentionally.

Z. Benenson et al. proposed the idea of using wearable sensors to measure different human body parameters e.g. body temperature, heart rate, blood pressure, perspiration, and brain impulses. The personality type of a particular person is determined from the sensed data. The person may be notified or advised based on his behavior in current context and how it may appear to his counterparts. The authors have presented a very novel idea but they did not discuss about any technical details. Such applications pose new technical requirements e.g. advances in wearability of tiny sensors, basic infrastructure support, routing algorithms, design of implantable devices that are powered through energy harvesting etc. Cooperation is challenge for enhancing interactions among people.

P. J. Marron et al. presented new category of applications called enhancing social interactions. The authors believe that such applications are far away to become reality. This is because of unavailability of the required technology and infrastructure. Research on such applications could possibly start after 5 years.

# PROPOSED WORK

This paper presents a basic infrastructure to build a Personality Wireless Sensor Network (PWSN) that consists of several sensor nodes. These sensor nodes communicate through wireless interface across human body called wireless body area network (WBAN). These sensor nodes are carried out by a person may be either embedded into clothes and other accessories or implantable in human body. The se sensor nodes are used to sense, measure and collect various parameters e.g. body temperature, blood pressure, perspiration, brain impulses, breathing activity etc. The collected data is transmitted periodically through radio transmitter to the cluster h ead (i.e. ACN node).

In addition to sensor nodes, each person carries a special node called ACN (Aggregator, Classifier & Notifier) node. This ACN node may be carried as a wrist -device with more resources (e.g. more battery power, higher processor and mem ory). ACN nodes are used to collect sensed data from sensor nodes. These ACN nodes carry out computation on received data locally and determine person's personality type. The person is notified about his behavior in current context and may generate some a dvices on how to behave with his counterpart. The received data is also transformed into information to determine person's psychological condition. The person's context-aware psychological condition is determined in different contexts by these ACN nodes.

The proposed infrastructure consists of following basic components:

**Sensor Nodes:** The functional block diagram of a typical sensor modified from (Akyildiz *et al.*, 2002) is shown in figure 1. It consists of five basic components i.e. sensing, computation, communication and battery. These sensor nodes are resource constrained in terms of communication, computation and battery power. The basic task of sensor nodes is to sense, measure or collect and transmit information to ACN nodes through wireless communi cation. The sensor nodes should be wearable or implantable.

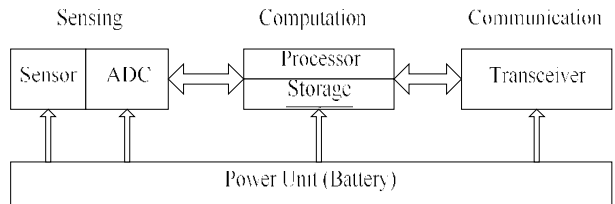

**Figure 1. Sensor node architecture for PWSN**

**ACN Nodes:** are rich in terms of resources i.e. higher communication, computation and energy. These nodes serve as cluster head and are used to aggregate data from sensor nodes. They are used to determine personality type and mood of owner in current context. They transform received data into information and determine psychological condition of the person. The owner is notified of its behavior and may receive some advices about how to app ear with his counterpart.

The PWSN consists of three parts:

**IntraPWSN (Intra Personality Wireless Sensor Network)**

IntraPWSN consists of sensor nodes and ACN nodes. They can communicate through wireless interface across human body. It is responsible for determining the mood and condition of the person. The data from different sensors is aggregated and processed into information. The person's personality type is determined based on this information. The person is notified about his behavior in current context and might be advised by the IntraPWSN to tune up his behavior after receiving information of counterpart. IntraPWSN is only concerned with a single person. The major responsibilities and infrastructure is presented in figure 2.

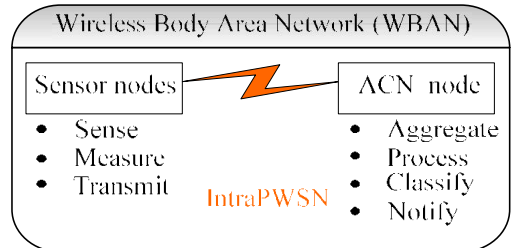

**Figure 2. Major responsibilities of nodes in IntraPWSN**

**InterPWSN (Inter Personality Wireless Sensor Networ k)**

InterPWSN is responsible when two or more persons with IntraPWSN meet together. The ACN nodes periodically wakeup and search for counterpart. Whenever there is some ACN node looking for it, they may ask each other through 'handshake'. They may exchange 'hello' messages. When both the ACN nodes have decided to communicate with each other, they first authenticate each other. The purpose of authentication is to avoid anonymous communication. They may share some key in order to authenticate each other. Once they have authenticated each other, they may issue a wake up call to sensor nodes to send of current context. The sensor nodes wake up and sense, collect and transmit current data to the ACN nodes. ACN nodes aggregate data from different sensors and proce ss it into useful information. ACN nodes determine personality type, mood and psychological condition of the owner. ACN nodes notify the owner with their personality type, mood and psychological condition in current context along with brief information of counterpart.

The owner can tune up his behavior or adjust itself with the current situation. If both the owners found themselves ready to interact they may allow their IntraPWSN to share their personal information like personality type, mood, psychologica l condition etc. After sharing personal information, now both the interacting parties know better each other. There is less probability so that they misunderstand or misinterpret each other. Because they interpret all the conversation in context of considering personality type of each other. Both IntraPWSN periodically updates each other with personal information and changing conditions.

On the other hand, they may leave each other by just sharing 'good bye' message which means they are not ready for this situation. The major responsibility of InterPWSN is to make successful interaction between different IntraPWSNs in order to avoid conflicting situation and help people understand each other. This is achieved through sharing personal information with each other and to let each other know in better way. Figure 3 shows the interaction of IntraPWSN 1 and IntraPWSN 2.

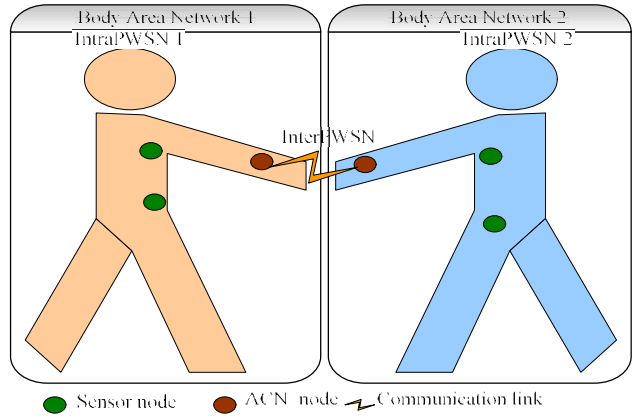
**Figure 3: Interaction of InraPWSNs**

Figure 4 shows the interaction of IntraPWSNs i.e. InterPWSN along with sensor and ACN nodes and their responsibilities.

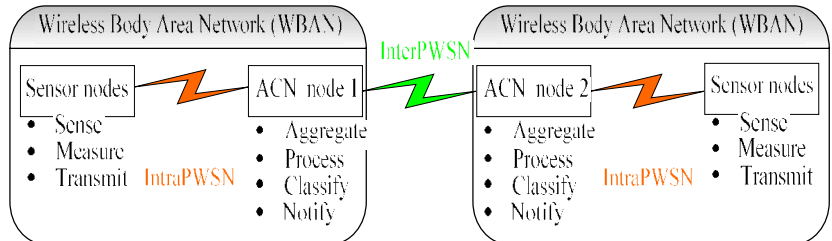
**Figure 4: Major responsibilities of nodes and their working in InterPWSN**

### A. *Remote PWSN Monitoring (Remote Personality Wireless Sensor Network Monitoring)*

A person can be monitored remotely about it's behavior in different contexts with different social groups.

PWSN can be seamlessly integrated either with 3G wideba nd cellular networks by utilizing CDMA2000 infrastructure or internet.

### 1. Integration of PWSN with 3G wideband cellular networks infrastructure

Significant numbers of remote patient monitoring applications are emerging in the field of telemedicine by employing existing networking infrastructure (Hu et al., 2003). To remotely monitor a person's behavior, sensed data is forwarded by ACNs to the cluster head through multihop communication. The cell phone or PDA serves as a cluster head. It aggregates data from ACNs and forward it to the base station i.e. Remote monitoring centre. It uses existing 3G wideband cellular infrastructure because ACNs cannot communicate directly with base station since their transmission range is within WPAN (Wireless Personal Area Ne twork). ACN may use multihop communication to transmit data to the other ACNs in case a cell phone or PDA is not directly reachable. Figure 5 shows the integration of PWSN with 3G cellular networks infrastructure for remote monitoring.

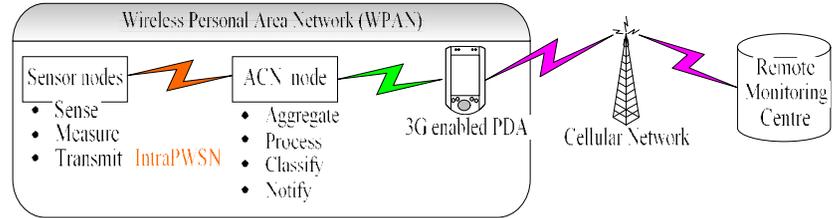

**Figure 5: Integration of PWSN with 3G cellular networks infrastructure for remote monitoring**

## 2. Integration of PWSN with internet infrastructure

PWSN can be integrated with existing internet infrastructure for remote monitoring of person's behavior. In this case, ACN nodes send their data to a PC, laptop or any other device having internet connectivity. The device having internet connectivity receives data from ACN nodes and forwards it to remote monitoring centre through internet. Figure 6 shows the integration of PWSN with internet infrastructure for remote monitoring.

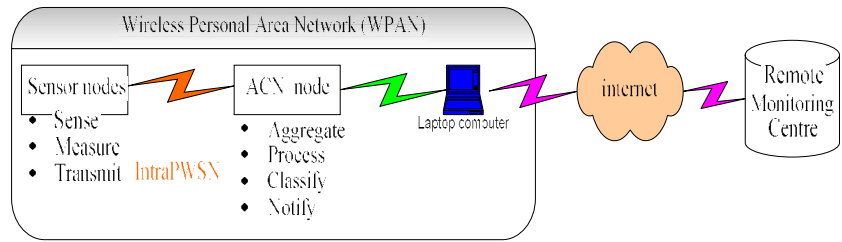

**Figure 6: Integration of PWSN with internet infrastructure**

In addition, serious patients (persons with serious psychological problems) can be monitored remotely using location sensors, motion or activity sensor, multimedia sensors (imaging, microphone and camera sensors).

## APPLICATION SCENARIO

There are numerous application scenarios whe re PWSN can be deployed but we are only considering an organization environment as case study.

An organization consists of number of individuals (hundreds or thousands) working in teams. The performance of working teams heavily depends on the attitude of t he individuals. This situation gets even more complicated in international working groups. Individuals working in teams need to understand each other in order to avoid potential conflicts.

PWSN can help individuals to understand each other by knowing the ir character to each other. It can also advise them, how to behave with their counterpart.

The behavior of employees can also be monitored in an organization by deploying PWSN. For example, a manager can remotely monitor behavior of his subordinates by d eploying remote

PWSN monitoring. We believe that an organization can get following benefits by deploying PWSN.

- The organization can have complete information of their employees regarding their personality type, mood and psychological classification. This can help them in identifying motivated, enthusiastic individuals. Organizations can help their employees in identifying and solving psychological problems when they are suffering from it.
- It will also help organizations to form project teams based on atti tude of individuals for better performance.
- It would be helpful for monitoring performance of individuals working in different teams.
- Improved social relationships among colleagues and service customers.

On the other hand, such infrastructure had few issue s that need to be addressed in future e.g. advances in wearability of tiny sensors, routing algorithms, design of implantable devices that are powered through energy harvesting, privacy issues, additional security infrastructure etc. Cooperation is challenge for enhancing interactions among people.

## CONCLUSION

This paper presents a basic infrastructure for personality wireless sensor networks. The idea is to use wearable or implantable tiny wireless sensors across human body to measure critical body parameters. The personal character or behavior (personality type, mood and psychological condition) is determined based on these parameters. The person is notified and advised of his own character in the current context. When two or more persons interact with each other, their personal character is exchanged to let them understand each other in better way. Their behavior can be monitored through PWSN remote monitoring.

We believe that PWSN can help peoples better understand each other and avoid potential conflicting situations. This may help them in tune up their behavior in different context.

## FUTURE DIRECTIONS

PWSN is an emerging application area of wireless sensor networks for human augmentation and enhancing social interactions. We envision that in near fu ture, wireless sensor networks would be integral part of our lives. Some of the future directions in the above categories other than presented in (Marron et al., 2006) are predicted here:

By sensing our circadian rhythms we can make more efficient use of o ur time by scheduling certain activities at certain times of the day. If we learn to listen to our bodies, we can work with these natural rhythms instead of fighting against them. We can determine from these natural circadian rhythms whether we are a "morn ing person" or a "night owl".

With the emergence of bio-sensors it would be possible to determine diseases a person might have e.g. diabetes, allergies etc. Environmental sensors being part of PWSN might send alerts/advises to take precautionary measures against different diseases e.g. pollen allergy etc.

## ACKNOWLEDGMENT

We are thankful to our friends and colleagues especially to Mr. Asfandyar khan, who have been very helpful in giving his suggestions and recommendations on remote PWSN monitoring.
## REFERENCES

Akyildiz, I. F., Su, W., Sankarasubramaniam, Y., Cayirci, E. 2002. Wireless sensor networks: a survey. Computer Networks: The International Journal of Computer and Telecommunications Networking, **v.38 n.4**, 393-422.

Benenson, Z., Günes, M., Wenig, M. 2006. PerSens: Personality Sensors (Poster), European *Workshop on Wireless Sensor Networks (EWSN'06), Sentient Future Competition*, Zurich, Switzerland.

Falikowski, A. Mastering Human Relations, 3rd edition, 2002, Pearson Education.

Hu, F., Kumar, S. 2003. QoS Considerations in Wireless Sensor Networks for Telemedicine. *Proc of SPIE ITCOM conference*, 217-227, Florida, USA.

Marron, P. J., Minder, D. 2006. Embedded WiSeNts Research Roadmap. *Technical report,* Germany.

Myers-Briggs (http://russellrowe.com/Myers-Briggs%20Typology%20System.htm)

Yick, J., Mukherjee, B., Ghosal, D. 2008. Wireless sensor network survey. *Computer Networks: The International Journal of Computer and Telecommunications Networking*,

Younis, M., Akkaya, K., Eltoweissy, M., Wadaa, A. 2004. On Handling QoS Traffic in Wireless Sensor Networks. *Proceedings of the Proceedings of the 37th Annual Hawaii International Conference on System Sciences (HICSS'04) -* **Track 9**, **V 9**, 90292.1.
9